\documentclass[12pt]{article}
\usepackage{amssymb,amsmath,amsthm,amsfonts}
\usepackage{pstricks,pspicture,graphicx}
\usepackage{epsfig,verbatim,alltt}
\usepackage{chicago}

\begin{document}

\title{Bayesian threshold selection for extremal models using measures of surprise}
\author{J. Lee\footnote{School of Computing \& Mathematical Sciences, Auckland University of Technology, Auckland, New Zealand.},\: 
 Y. Fan\footnote{School of Mathematics and Statistics, University of New South Wales, Sydney, Australia.}\: 
 and S. A.  Sisson$^\dagger$\footnote{Communicating Author: Scott.Sisson@unsw.edu.au. Phone: +612 9385 7027. Fax: +612 9385 7123.}}

 \maketitle

\begin{abstract}

Statistical extreme value theory is concerned with the use of asymptotically motivated models to describe the extreme values of a process.
A number of commonly used models are  valid for observed data that exceed some high threshold. However, in practice a suitable threshold is unknown and must be determined for each analysis.
While there are many threshold selection methods for univariate extremes, there are relatively few that can be applied in the multivariate setting.
In addition, there are only a few Bayesian-based methods, which are naturally attractive in the modelling of extremes due to data scarcity.
The use of Bayesian measures of surprise to determine suitable thresholds for extreme value
models is proposed.
Such measures quantify the level of support for the proposed extremal model and threshold, without the need to specify any model alternatives. This approach is easily implemented for both univariate and multivariate extremes.
\\

\end{abstract}

\section{Introduction}
\label{section:intro}

Extreme value theory is often used for the modelling of rare events in many applied areas, including finance \shortcite{embrechts+km03}, engineering \shortcite{castillo+hbs04} and the environmental sciences \cite{coles01}. Commonly, a mathematically derived parametric extreme value model is used to describe the tail of the data generation process above some high threshold.

In the univariate case, the generalised Pareto distribution provides a suitable model for the analysis of threshold exceedances, under mild conditions \cite{pickands75,balkema+d74,davison+s90}. Specifically, if $X_1, X_2, \ldots \in\mathbb{R}$ denote a sequence of independent and identically distributed random variables, then the asymptotic distribution of the exceedances, $Y=X-u|X>u$, of some high threshold $u$ is given by
\begin{equation}
\label{eqn:GPD}
	F(y|\xi,\sigma,u) = 1-\left[1+\dfrac{\xi (y-u)}{\sigma}\right]_+^{-1/\xi},
\end{equation}
where $[a]_+=\max\{0,a\}$, and $\sigma>0$ and $-\infty<\xi<\infty$ denote scale and shape parameters. The generalised Pareto model (\ref{eqn:GPD}) holds as $u\rightarrow\infty$, and so in practice a suitable choice of threshold is the smallest value, $u$, such that $F$ approximates the tail of the observed data sufficiently well.

In the multivariate setting, a standard representation is given in terms of a limiting Poisson process \cite{deHaan85,resnick87}. If $Z_1,\ldots, Z_n\in\mathbb{R}^d$ are an independent and identically distributed sequence of random vectors with unit Fr\'echet margins (i.e. with distribution function $\exp(-1/z)$, for $z>0$), then the sequence of point processes $P_n=\{Z_i/n:i=1,\ldots n\}$ on $[0,\infty)^d$ converges to a non-homogeneous Poisson process $P_n\rightarrow P$ on $[0,\infty)^d\backslash \{0\}$ as $n\rightarrow\infty$ \cite{deHaan85}. The intensity function of $P$ has the form
\begin{equation} 
\label{eq:intensity}  
	\nu(dz)= \dfrac{dr}{r^2}H(dw),
\end{equation}
where $(r, w)$ denotes the pseudopolar co-ordinates $r=\frac{1}{d}\sum_{i=1}^d Z^i$ and $w=Z/r$ 
(where $Z=(Z^1,\ldots,Z^d)\in\mathbb{R}^d$), 
and $H$ is a measure function defined on the unit simplex which represents the multivariate dependence structure.
As with the univariate case the above Poisson process holds asymptotically, and so in practice (that is, for finite $n$) it is assumed to hold approximately on regions bounded away from the origin. In this case, the Poisson process with intensity function (\ref{eq:intensity}) may be fitted to those observations $(r, w)|r>r_0$ for which $r$ exceeds some high threshold, $r_0$. As before, a suitable choice of threshold is the smallest value, $r_0$, such that the above Poisson process approximates the multivariate tails of the observed data sufficiently well (e.g. \citeNP{coles+t91})\textcolor{red}{.}

In both univariate and multivariate settings, the choice of a suitable threshold ($u$ or $r_0$) is problem dependent. As such, a number of approaches have been proposed, primarily for the univariate case, that either offer diagnostics for threshold choice or estimate the threshold as part of the model fitting procedure. A comprehensive review of these methods for the choice of $u$ is given by \citeN{scarrott+m12}, who loosely characterise the techniques into several categories. 

Classical fixed threshold approaches use graphical or other diagnostics to make an assessment of the model fit, in order to make an a priori threshold choice. These include e.g. mean residual life plots, threshold stability plots, Hill plots and general distribution fit diagnostics (e.g. \shortciteNP{davison+s90,beirlant+vt96,dupuis98,drees+dr00,coles01,choulakian+s01}). Disadvantages to these approaches are that graphical diagnostics are sometimes difficult to correctly interpret, and that the uncertainty associated with the threshold, $u$, is not well accounted for in the frequentist framework although see \citeNP{CabrasCastellanos2009b} who develop a Bayesian mean residual life plot). 
Some methods that have been proposed to informally overcome these problems include tail fraction estimation \shortcite{drees+dr00,dekkers93,feuerverger+h99,goegebeur+bw08} and resampling-based approaches \shortcite{danielsson+dpv01,ferreira+dp03,beirlant+vt96,drees+k98}.

Rather than making an a priori threshold choice, several Bayesian mixture models have been proposed which treat the threshold as an unknown parameter to be estimated. The mixture components themselves correspond to the generalised Pareto model above $u$, and parametric or semi-/non-parametric estimators of the bulk of the distribution below the threshold \shortciteNP{frigessi+hr03,behrens+hg04,tancredi+ao06,cabras+m07,cabras+c11,macDonald+sldrr11}). These approaches are attractive as they both incorporate threshold uncertainty in the analysis, and also remove the need to make a subjective decision on the value of a fixed threshold. Disadvantages of these approaches are the need to correctly balance the relative influence of the bulk and Pareto mixture components so that neither dominate \shortcite{macDonald+sl11}, and that it would appear difficult to extend them to the multivariate setting.

All of the above approaches concern the Pareto threshold, $u$, for univariate extremes. There are virtually no diagnostics to determine the threshold, $r_0$, for multivariate extremes models. However, noting that the intensity measure (\ref{eq:intensity}) is expected to factorise into independent components involving angular ($w$) and radial ($r$) components when $r>r_0$, in principle diagnostics may be constructed by determining the smallest value of $r_0$, such that $r|r>r_0$ and $w|r>r_0$ exhibit independence. Intuitively, this is easiest to achieve for bivariate extremes, so that both $w$ and $r$ are univariate, whereby empirical histograms of $w|r>r_0$ should visually retain the same shape for $r>r_0$ (e.g. \shortciteNP{joe+sw92,coles+t94}).

In this article we propose a new Bayesian diagnostic for threshold choice for extremal models based on the idea of ``surprise''  \cite{meng94,bayarri+m03,bayarri+b98,cabras+m07}.  Measures of surprise quantify the degree of incompatibility of observed data with a given model, commonly through various (Bayesian) predictive $p$-values using appropriate test statistics, but without any reference to alternative models. 
In terms of threshold identification, such measures would enumerate the extent to which observed data exceeding a candidate threshold ($u$ or $r_0$) are compatible with the asymptotic  Pareto or point process model. The smallest threshold values that are not incompatible with the data are then natural candidates for the selected threshold.

Unlike many existing threshold choice methods, as predictive $p$-values have a natural scale, these measures of surprise allow direct comparison of competing threshold candidates (which have different amounts of data exceeding the threshold), as they do not require any modelling of data below the threshold. Also unlike almost all existing methods, by construction, this approach is equally applicable to both univariate and multivariate extremal models.
Ultimately, the proposed surprise-based approach will select  a final fixed threshold, $u$ or $r_0$, for use in a subsequent analysis. As a result, threshold uncertainty is not directly incorporated into this final analysis. However, the threshold selection procedure itself is fully Bayesian, and the final choice of threshold can be made within a full Bayesian decision-theoretic framework.

The remainder of this article is organised as follows: Section \ref{sec:ms} provides a brief introduction to measures of surprise and various forms of predictive $p$-value, before describing the proposed threshold selection procedure. The performance of this approach is evaluated through several simulated examples in Section \ref{sec:sim}, both univariate and multivariate, and compared to existing approaches for threshold choice. In Section \ref{sec:application} we apply our procedure to several real examples that have been previously analysed in the extremes literature. Finally, we conclude with a discussion.

\section{Using surprise for threshold selection}
\label{sec:ms}

\subsection{Surprise and posterior predictive $p$-values}%
\label{section:2.1}

Suppose that we are interested in determining whether observed data, $y_{obs}=(y_{obs,1},\dots,y_{obs,n})$, are consistent with some null hypothesis $H_0: y_{obs}\sim f(y|\theta)$, where $f$ is a density function with parameters $\theta\in\Theta$. In the present setting, $f(y|\theta)$ corresponds to the Pareto density function (\ref{eqn:GPD}) in the univariate case, or a density corresponding to the measure (\ref{eq:intensity}) for multivariate data (e.g. see Section \ref{sec:pval}). In the Bayesian setting, hypothesis testing is most commonly performed through computation of Bayes Factors (e.g. \shortciteNP{kass+r95,han+c01}). However the need to specify an alternative model is problematic in the identification of a threshold for an extremal model. For example, given a threshold, do the data more likely come from a Pareto distribution or a specific other distribution? Clearly, there is no generic alternative model that would adequately describe deviations from the Pareto in all cases.

Previous Bayesian approaches to univariate threshold selection have circumvented this problem by treating the threshold as an unknown random variable, and modelling the bulk of the data below the threshold by a semi- or non-parametric estimate such as a mixture distribution (\shortciteNP{frigessi+hr03,behrens+hg04,tancredi+ao06,cabras+m07,cabras+c11,macDonald+sldrr11}). 
These approaches are able to rely on a single model hypothesis, $H_0$, and then probabilistically identify which portion of the upper tail of the data best satisfy this hypothesis.
However, they can be sensitive to the form of semi-/non-parametric estimates used, and it is not obvious how to practically
 extend the sub-threshold modelling to bivariate or higher dimensional models.
  As an alternative, we turn to the Bayesian idea of ``surprise.''

Measures of surprise are designed to quantify the degree of consistency of the data with $H_0$, without specification of an alternative hypothesis. 
There is a huge literature concerning different quantifications of this concept \shortcite{weaver48,guttman67,box80,berger85,good88,meng94,evans97,bayarri+b98} -- see e.g. \citeN{bayarri+b98} for a critical review.
In classical statistics, a natural measure of surprise is the $p$-value 
\begin{equation}
\label{eqn:pvalue}
	p=\mbox{Pr}_{f}(T(y) \geq T(y_{obs})),
\end{equation}
computed with respect to the null distribution $f$, test statistic $T$ and observed data $y_{obs}$. 
The $p$-value measures how likely it is to observe data more extreme than the observed data, $y_{obs}$, if the null model $f(y|\theta)$ is correct.

In the Bayesian framework,  the Bayesian $p$-value is obtained by considering various choices of the probability distribution used to compute (\ref{eqn:pvalue}), in the sense of integrating out the unknown parameters, $\theta$. These include the prior predictive distribution \cite{box80},
\[
m(y) = \int f(y|\theta)\pi(\theta)d\theta,
\quad\mbox{ so that }\quad
p_m=\mbox{Pr}_m(T(y)\geq T(y_{obs})),
\]
where $\pi(\theta)$ and $f(y|\theta)$ denote the prior and likelihood, respectively, and the posterior predictive distribution (e.g. \citeNP{guttman67,Rubin1984}),
\begin{equation}
\label{eqn:pm0}
m_0(y|y_{obs}) = \int f(y|\theta)\pi(\theta|y_{obs})d\theta,
\quad\mbox{ with }\quad
p_{m_0}=\mbox{Pr}_{m_0}(T(y)\geq T(y_{obs})),
\end{equation}
where $\pi(\theta|y_{obs})\propto f(y_{obs}|\theta)\pi(\theta)$ is the posterior distribution for $\theta$ given $y_{obs}$. 
Computation for $p_m$ and $p_{m_0}$ is straightforward: samples $\theta^{(i)}\sim\pi(\theta)$ drawn from the prior (or $\theta^{(i)}\sim\pi(\theta|y_{obs})$ drawn from the posterior, in the case of $p_{m_0}$) are used to  produce samples from $m(y)$ (or $m(y|y_{obs})$) under the model, $f(y|\theta)$. The predictive $p$-values may then be estimated by the proportion of times that $T(y)\geq T(y_{obs})$.
However, while simple, the prior predictive $p$-value, $p_m$, can perform unsatisfactorily if $\pi(\theta)$ is poorly elicited or improper.  Similarly, the posterior-predictive $p$-value, $p_{m_0}$, is easy to compute using posterior samples obtained from standard Monte Carlo algorithms, however  it can be criticised by its double usage of the likelihood within the distribution $m_0(y|y_{obs})$ (e.g. \citeNP{bayarri+b98}).

To circumvent issues involving the impropriety of $\pi(\theta)$  in $p_m$, and the double usage of the observed data in $p_{m_0}$, \citeN{bayarri+b98} proposed the partial posterior predictive $p$-value, defined as
\begin{equation}
\label{eqn:pmstar}
m^*(t) = \int f(t|\theta)\pi(\theta|y_{obs}\backslash t_{obs})d\theta,
\quad\mbox{ and }\quad
p_{m^*}=\mbox{Pr}_{m^*}(t\geq t_{obs}).
\end{equation}
Here, $t=T(y)$ corresponds to the test statistic of interest, $f(t|\theta)$ is the (known) density function of the test statistic, and $\pi(\theta|y_{obs}\backslash t_{obs})$ is the (partial) posterior distribution of $\theta$ obtained using the observed data, $y_{obs}$, but excluding the test datum $t_{obs}=T(y_{obs})$. In this manner, the full dataset, $y_{obs}$, is only used once within $m^*(t)$. 
Posterior simulation from the partial posterior, $\pi(\theta|y_{obs}\backslash t_{obs})\propto f(y_{obs}|\theta)\pi(\theta)/f(t_{obs}|\theta),$ is available using regular posterior simulation algorithms.

In the classical setting, 
the $p$-value is uniformly distributed on (0,1) if the fitted model is correct. Using the weaker property
of the asymptotic uniformity of the $p$-value \shortcite{RobinVanderVaartVentura2000}, 
\citeN{meng94} showed that the expected value of the posterior predictive $p$-value is 0.5 under the null (see also \citeNP{gelman13}). Hence,  $p$-values close to 0 or 1 (depending on the test statistic, $T(y)$ -- see Section \ref{sec:test}) can be safely be interpreted as indicating incompatibility of the observed data, $y_{obs}$, with the null model. Conversely, $p$-values close to 0.5 indicate a lack of evidence against the hypothesised model, although as argued by \citeN{BayarriCastellanos2007}, even a $p$-value of 0.4 can not naively be interpreted as compatibility with the null model in all problems.

\subsection{Predictive $p$-values for threshold models}%
\label{sec:pval}

For the purposes of threshold estimation for threshold exceedance models using measures of surprise, we will make use of the full and partial posterior predictive $p$-values, $p_{m_0}$ (\ref{eqn:pm0}) and $p_{m^*}$ (\ref{eqn:pmstar}). 
In the univariate case, for those observations $Y=X-u|X>u$ that exceed the threshold $u$, the model $f(y|\theta)$ is the density function derived from the Pareto distribution (\ref{eqn:GPD}).
Note that the null hypothesis is conditional on a fixed threshold  $u$ (and similarly for the multivariate setting).
In the multivariate case, following transformation to unit Fr\'echet margins and the pseudo-polar co-ordinates $(r,w)$, 
when the radial component $r$ exceeds the marginal threshold, $r_0$, 
the Poisson process intensity function (\ref{eq:intensity}) factorises into separate terms involving $r$ and $w$. As a result, the model $f(y|\theta)$ reduces to $f(y|\theta)\propto dH(w)$ (e.g. \citeNP{coles+t94}).
However, the measure function $H(w)$
does not admit any universal closed form representation in general. Instead, various forms of parametric families have been derived for the corresponding 
function, $h(w)= dH(w)$ (where available), that satisfy its definition on the unit simplex, $S_d=\{w\in\mathbb{R}^d_+: \sum_{i=1}^dw_i=1\}$, and the marginal normalisation constraints $\int_{S_d}w_idH(w)=1$, for $i=1,\ldots,d$ (e.g. \citeNP{coles+t94}).

A common, and simple choice for the dependence measure function for bivariate extremes, is the bilogistic model \shortcite{joe+sw92}, with
\begin{equation} 
	\label{eq:bilogistic}
	h(w|\alpha,\beta) = \dfrac{(1-\alpha)(1-\gamma)\gamma^{1-\alpha}}{(1-w)w^2 [\alpha(1-\gamma)+\beta\gamma]},
\end{equation}
where $0<\alpha,\beta<1$ and $\gamma$ is the root of
$(1-\alpha)(1-w)(1-\gamma)^{\beta}-(1-\beta)w\gamma^{\alpha}=0$.
Here, $(\alpha^{-1}-\beta^{-1})/2$ and $(\alpha^{-1}+\beta^{-1})/2$ are treated as measures of asymmetry and dependence strength, respectively. When $\alpha=\beta$, the model reduces to the symmetric bivariate logistic model.
Other closed form model parameterisations for the dependence measure, $H(w)$ (and $h(w)$), are available for both bivariate and multivariate extremes (see e.g.  \citeNP{kotz+n02} for an overview).

However, it is generally accepted that most parametric models for $h(w)$ are insufficiently flexible for accurate modelling of tail behaviour in more than two dimensions, whereas non-parametric approaches have only been developed for two- or three-dimensional data. \citeN{BoldiDavison2007} proposed the use of a mixture of dirichlet distributions as a semi-parametric model to adequately model higher dimensional problems. In this setting,
\begin{equation} 
	\label{eq:dirichlet}
	h_D(w|\mu,\lambda) = \sum_{i=1}^I \lambda_i f_D(w|\mu^i),
\end{equation} 
subject to the necessary marginal normalisation constraints, 
where $f_D(w|\mu^i)$ is the dirichlet density function with parameter vector $\mu^i=(\mu^i_1,\dots,\mu^i_d)$, $\lambda=(\lambda_1,\ldots,\lambda_I)$ are positive weights such that $\sum_{i=1}^I \lambda_i=1$, and $\mu=(\mu^1,\ldots,\mu^I)$. An efficient algorithm (with code) to sample from the posterior distribution, $\pi(\mu, \lambda,I|w)\propto h_D(w|\mu,\lambda)\pi(\mu,\lambda)\pi(I)$, using Bayesian model averaging, is described in \citeN{sabourin+n13}, although for the simulations in this article we fixed $I$ at the largest value such that no components were empty, following asymptotic arguments by \citeN{RousseauMengersen2011}.

\subsection{The test statistic, $T(y)$}%
\label{sec:test}

The test statistic, $T(y)$, is chosen to investigate the compatibility of the observed data, $y_{obs}$, with the model -- here, a Pareto model  or Poisson process -- where typically, large values of $T$ indicate less compatibility (e.g. \citeNP{bayarri+b98}). Selection of $T(y)$ is problem specific, and a clear choice is usually only available for very simple problems.
The maximum, minimum and mean of the data have been commonly used for a range of problems  (e.g. \citeNP{bayarri+b98,bayarri+m03}).

In the case of threshold exceedance models for extremes, we consider two choices for $T(y).$ 
The first is
the reciprocal likelihood function, $T(y)=1/f(y|\theta)$, which provides a measure of the overall adherence of the data to the model. 
This statistic has the advantage that it is easily defined and evaluated for both univariate and multivariate models for extremes, but has the disadvantage that it is difficult to compute its distribution, $f(t|\theta)$. As such it is only easily implemented for the posterior predictive  $p$-value, $p_{m_0}$.
Here, a small $p$-value corresponds to evidence against the Pareto or point process model, for the given threshold $u$ or $r_0$.

The second statistic is $T(y)=q_j(y)$, the $j$-th empirical quantile of $y$. This statistic permits a more precisely located examination of data and model compatibility, which may be more suitable in circumstances where fidelity at particular quantiles (such as near the maximum) is more important than than overall model fit. An advantage of this statistic is that it can be used with both the posterior predictive, and partial posterior predictive $p$-values, $p_{m_0}$ and $p_{m^*}$. In the latter case, the distribution $f(t|\theta)$ is given by the distribution of order statistics (e.g. \citeNP{DavidNagaraja2003}), so that for a dataset $y=(y_1,\ldots,y_n)$, with $y_{(1)}<y_{(2)}<\cdots <y_{(n)}$, we have
\[
	f(t|\theta) =
	f(y_{(j)}|\theta) =
	\dfrac{n!}{(j-1)!(n-j)!} f(y_{(j)}|\theta)F(y_{(j)}|\theta)^{j-1}[1-F(y_{(j)}|\theta)]^{n-j} .
\]
Here $F$ and $f$ correspond to the distribution and density function of the generalised Pareto distribution (\ref{eqn:GPD}). However, an obvious disadvantage of this statistic is that it is only defined for univariate analyses i.e. determination of $u$. 
Here, as the magnitude of $T(y)$ does not directly quantify compatibility with the extremal model, either small or large $p$-values indicate evidence against the Pareto model.

\subsection{Method summary}%

A brief summary of the proposed approach for threshold selection is given as follows.
\begin{enumerate}
\item[1)]
Define a series of candidate thresholds, $v_1>\ldots>v_m$, where $v=u$ for univariate models and $v=r_0$ for multivariate models. 
\item[2)] For $i=1,\ldots,m$:
\begin{enumerate}
\item[(a)] 
Obtain draws from the posterior distribution $\pi(\theta|y_{obs}(v_i))$ or partial posterior distribution $\pi(\theta|y_{obs}(v_i)\backslash t_{obs}(v_i))$, where $y_{obs}(v_i)=\{y_{obs,k}:y_{obs,k}>v_i\}$ is the subset of elements of $y_{obs}$ that exceed the threshold, $v_i$, and $t_{obs}(v_i)=T(y_{obs}(v_i))$.
\item[(b)] Generate samples from $m_0(y(v_i)|y_{obs}(v_i))$ or $m^*(t(v_i))$, given by (\ref{eqn:pm0}) and (\ref{eqn:pmstar}) based on the posterior from (a), and compute threshold-dependent empirical estimates of $p_{m_0}$ or $p_{m^*}$.
\end{enumerate}
\item[3)] Construct a plot of threshold versus $p$-values, and determine the most suitable threshold.
\end{enumerate}

An advantage in examining candidate thresholds from the largest ($v_1$) to the smallest ($v_m$) is that computational efficiencies can be achieved in sampling from $m$ posterior distributions, $\pi(\theta|y_{obs}(v_i))$ or $\pi(\theta|y_{obs}(v_i)\backslash t_{obs}(v_i))$. That is, there are likely to be strong similarities between two successive distributions e.g. $\pi(\theta|y_{obs}(v_i))$ based on $v_i$ and then $v_{i+1}$, as there is only a small difference in the data used in their construction. As such, it is natural to adopt sequential sampling techniques, such as sequential Monte Carlo (e.g. \shortciteNP{delmoral+dj06,DoucetFreitasGordon2001}), to perform posterior simulation for all $m$ models in a single algorithm implementation, rather than run (say) $m$ independent MCMC samplers. Efficient data assimilation techniques, whereby each distribution in a sequence uses increasingly larger amounts of data, are commonplace in sequential Monte Carlo algorithms (e.g. \shortciteNP{DoucetFreitasGordon2001}).

There are a number of ways to select an appropriate threshold sequence, $v_1>\ldots>v_m$. The simplest is to specify $m$ equally spaced values between notional maximum and minimum thresholds, $v_1$ and $v_m$. An alternative is to first specify $v_1$, and then repeatedly lower the threshold by including fixed amounts of data, until it is clear from the measure of surprise that a lower bound has been reached. More adaptive approaches to selecting the thresholds can be constructed by observing that $p$-values for lower thresholds only require computing if there is a significant change in their values as more data is included. This allows for strategies based on the monitoring of e.g. maximum likelihood estimates, or the effective sample size (a measure of the variability of importance weights) within sequential sampling algorithms, as the threshold decreases, or where the threshold is decreased until a pre-determined change in these quantities occurs. 
For the simulations in this paper, we implement the simplest approach of $m$ equally spaced values.

In the univariate case, a plot of threshold ($u$) versus surprise ($p$-value) is sufficient to evaluate the compatibility of data and model for each threshold. In the multivariate case, a similar plot of $r_0$ versus $p$-value can be produced. However, in the multivariate setting we are able to examine more than just whether the data and model are compatible at the candidate threshold $r_0$. Here, the intensity function of the non-homogeneous Poisson process (\ref{eq:intensity}) factorises if $r_0$ is sufficiently large, implying that the distribution of $w|r>r'_0$ does not change for $r'_0>r_0$. 
This condition can be evaluated directly by measuring surprise at thresholds $r'_0>r_0$ based on the posterior generated at the threshold $r_0$, i.e. $\pi(\theta|y_{obs}(r_0))$ or $\pi(\theta|y_{obs}(r_0)\backslash t_{obs}(r_0))$. If the required factorisation of (\ref{eq:intensity}) exists, then the resulting $p$-values will not change as $r'_0$ increases.
In this manner we are able to avoid potentially erroneous conclusions where the $p$-value at a given threshold does not show incompatibility between data and model, but where the dependence assumption is not satisfied. Some illustrations of this can be seen in Figure \ref{fig:aq} (a) and (f).

Procedurally, constructing the additional $p$-values corresponds to including the following component in step 2) above:
\begin{enumerate}
\item[2) (c)]  For $j=(i-1),\ldots,2,1$:
Generate samples from $m_0(y(v_j)|y^*_{obs}(v_i))$ or $m^*(t(v_j))$, given by (\ref{eqn:pm0}) and (\ref{eqn:pmstar}), and, based on the posterior from (a),  compute threshold-dependent empirical estimates of $p_{m_0}$ or $p_{m^*}$.
\end{enumerate}

In this setting, the resulting plot of threshold versus surprise will consist of one line of $p$-values, as generated in steps 2) (b) and (c), for each candidate threshold $v_i$. Examples of these plots are illustrated in Figures \ref{fig:simulated-multivariate}, \ref{fig:ocean} and \ref{fig:aq}

\section{Simulated data analyses}
\label{sec:sim}

We first examine the utility of measures of surprise for threshold choice for simulated data in both the univariate and multivariate setting, where the true threshold is known. In the univariate setting we are also able to compare the use of surprise against several standard approaches for threshold detection, both classical and Bayesian.

\subsection{Univariate analyses}%

\begin{figure}
\centering
\includegraphics[width=17cm]{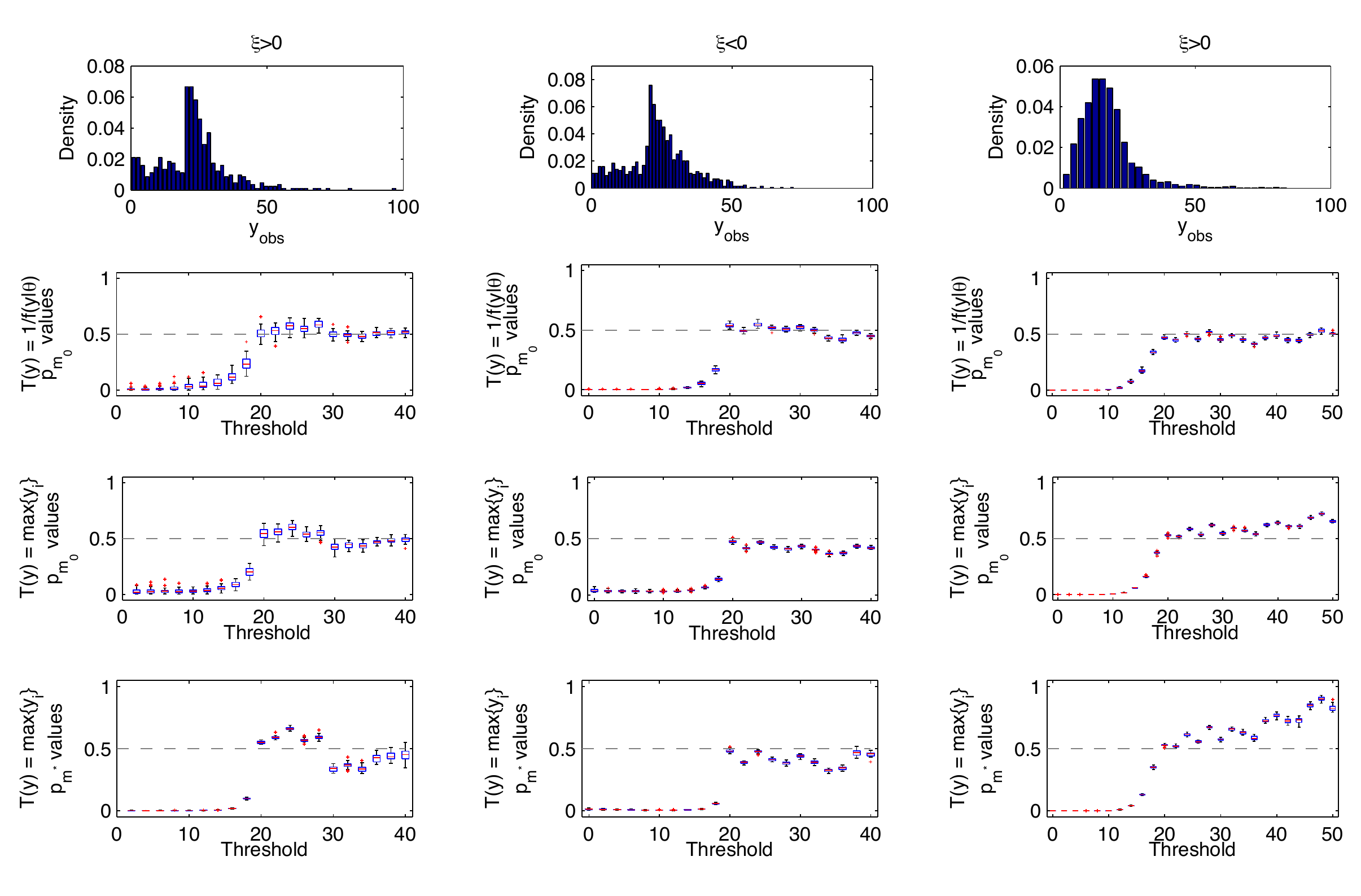} 
\caption{
\label{fig:univariate-simulated}
\small
Histograms and measures of surprise for each of the  three univariate datasets, each with a known threshold of $u=20$. 
Top panels show histograms of each dataset. The second and third rows respectively illustrate posterior predictive $p$-values, $p_{m_0}$, versus threshold, using the test statistics $T(y)=1/f(y|\theta)$ and $T(y)=\max\{y_i\}$. Bottom panels show partial posterior predictive $p$-values, $p_{m^*}$, using the statistic $T(y)=\max\{y_i\}$.
}  
\end{figure}

We generate three observed datasets, $y_{obs}=(y_{obs,1},\ldots,y_{obs,n})$,  from the mixture distributions:
\begin{itemize}
\item $\xi>0$: $y_{obs,i} \sim 0.3 U{(0,20)} + 0.7 f(\xi=0.2,\sigma=8,u=20)$, $n=500$
\item $\xi<0$: $y_{obs,i} \sim 0.3 U(0,20) + 0.7 f(\xi=-0.1,\sigma=10,u=20)$, $n=1,000$
\item $\xi>0$: $y_{obs,i} \sim 0.7 \mbox{Gamma}(3,8,u=20) + 0.3 f(\xi=0.4,\sigma=6.0974,u=20)$, $n=2400$
\end{itemize}
where $f$ denotes the density function of the Pareto distribution (\ref{eqn:GPD}), and
$U(\alpha,\beta)$ is the uniform distribution between $\alpha$ and $\beta$. $\mbox{Gamma}(\alpha,\beta,u)$ denotes the density function of a $\mbox{Gamma}(\alpha,\beta)$ distribution, with upper truncation at $u$.
Here, the true thresholds are known as $u=20$ in each case. 
The observed datasets are illustrated in Figure \ref{fig:univariate-simulated} (top row), and exhibit varying degrees of 
 Pareto tail behaviour.  Ten observations exceed 100 for the third dataset. 
The abrupt step change in the histograms of the first two datasets shows the true threshold very clearly. Whereas for the third dataset, which has Pareto distribution parameters chosen to make the density continuous at $u=20$, 
it is less clear, and is perhaps closer to a real world situation.

\begin{figure}
\centering
\includegraphics[width=16cm]{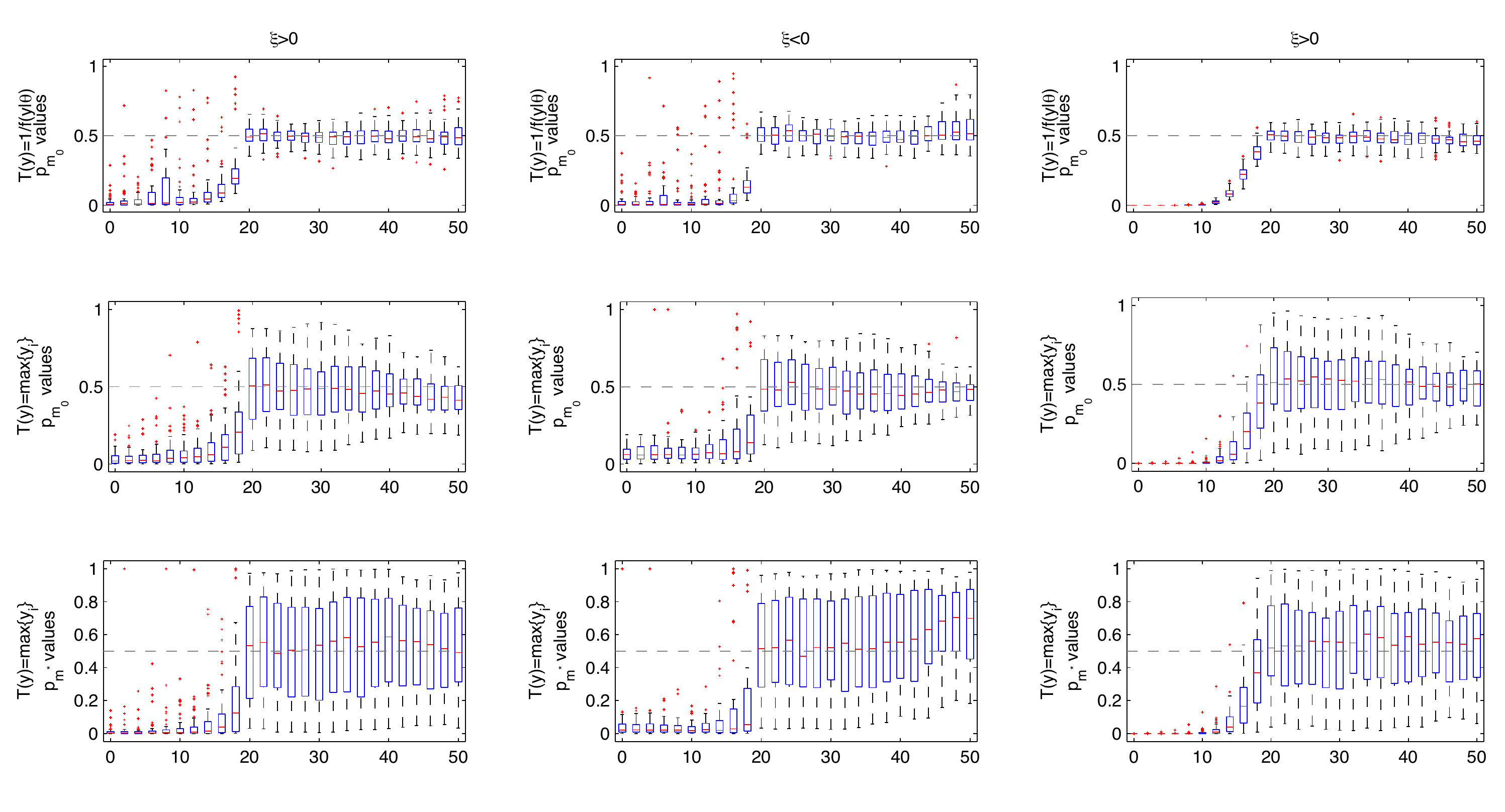} 
\caption{
\label{Fig:univariate-repeated}
\small
Dataset-induced variability in the measures of surprise for each of the  three univariate datasets, replicated over 50 observed datasets, $y_{obs}$. 
The top and middle rows respectively illustrate posterior predictive $p$-values, $p_{m_0}$, versus threshold, using the test statistics $T(y)=1/f(y|\theta)$ and $T(y)=\max\{y_i\}$. Bottom panels show partial posterior predictive $p$-values, $p_{m^*}$, using the statistic $T(y)=\max\{y_i\}$. 
}  
\end{figure}

\begin{figure}
\centering
\includegraphics[width=15cm]{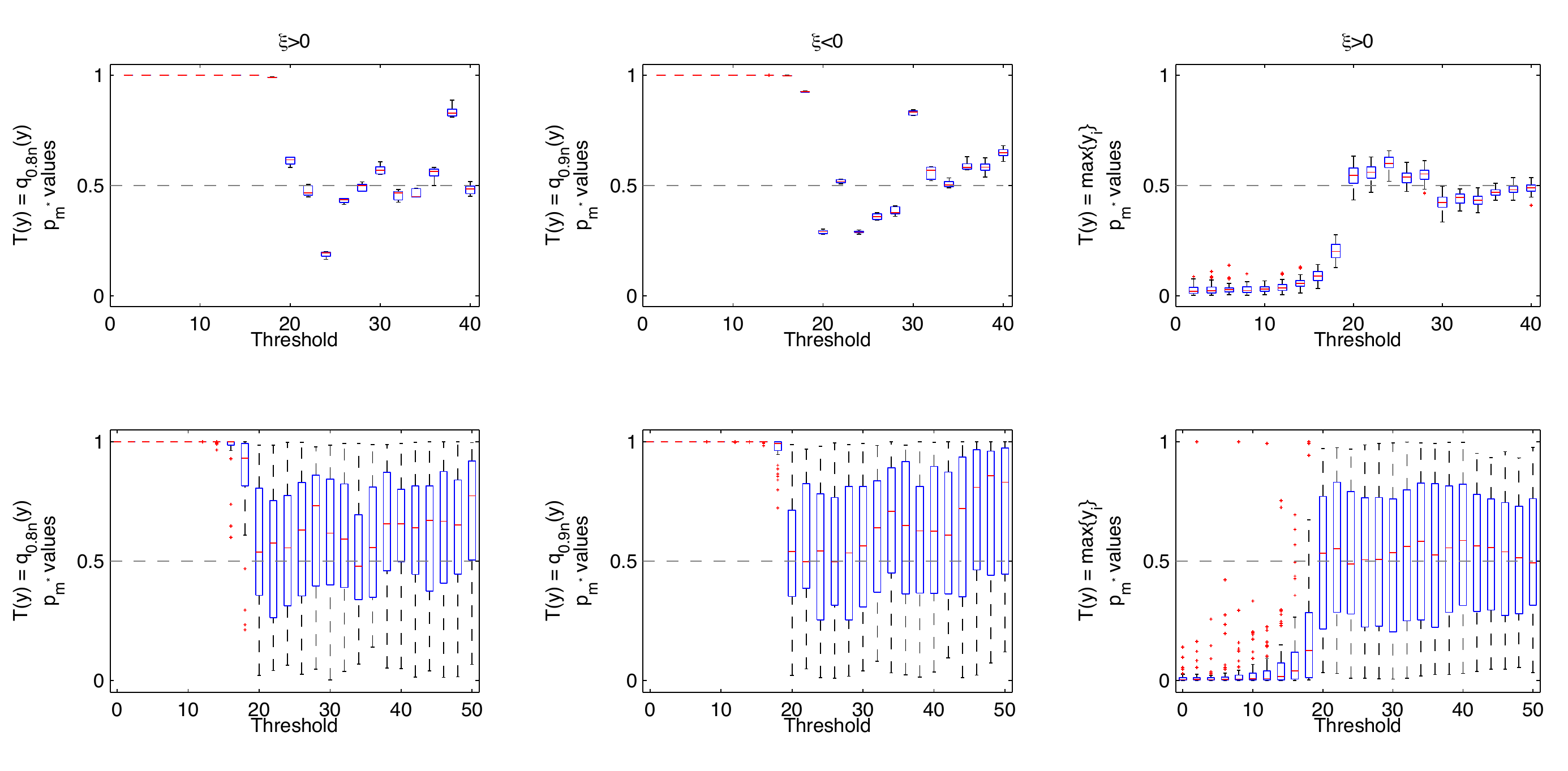} 
\caption{
\label{Fig:test_quantiles}
\small
Partial posterior predictive $p$-values for the first dataset with $\xi>0$, using the empirical quantile, $T(y)=q_j(y)$, as the test statistic. Left to right, panels show the measure of surprise versus threshold for $j=0.8n, 0.9n$ and $n$ (i.e. $T(y)=\max\{y_i\}$).
Top panels indicate  variability of $p_{m^*}$  for a single observed dataset, $y_{obs}$; bottom panels indicate variability over 50 replicate datasets.
}  
\end{figure}

For each observed dataset, $y_{obs}$, we computed various measures of surprise for each of the candidate thresholds $u=2, 4, 6, \ldots, 40$, based on 9,000 MCMC samples from the posterior at each threshold (with another 1,000 samples discarded as burn-in, based on convergence checks). We used the Jeffrey's prior for $\pi(\xi,\sigma)$ \cite{CastellanosCabras2007}, given by 
\[
	p(\xi,\sigma) \propto \sigma^{-1} (1+\xi)^{-1} (1+2\xi)^{-1/2},  \qquad\mbox{for } \xi>-0.5, \sigma>0.
\]
All 9,000 posterior samples were used to compute $p_{m_0}$ and $p_{m^*}$.

The second row of Figure \ref{fig:univariate-simulated} illustrates box-plots of 30 replications of the posterior predictive $p$-value, $p_{m_0}$, based on the reciprocal likelihood function, $T(y)=1/f(y|\theta)$, as the test statistic. While in practice, only a single replicate will be computed at each threshold, the boxplots highlight the low Monte Carlo variability of the $p$-value estimate based on $n=500$--$2,400$ observations and $9,000$ posterior samples, indicating rapid sampler convergence and good mixing.
For each observed dataset, $p_{m_0}$ is approximately constant at around 0.5 (indicated by the horizontal dashed line) for threshold values of $u_0$ at or above the true threshold. This is the expected outcome if the Pareto model is true \cite{meng94}. As the threshold moves below the true threshold in each case, the behaviour of the predictive $p$-value abruptly changes and gradually becomes small, indicating that the observed data, $y_{obs}$, is becoming increasingly unlikely under the model as non-Pareto-distributed  data are included in the analysis.
Note that reducing the threshold until $p_{m_0}<0.05$ (say) to identify the most suitable threshold, is not the correct way to interpret this plot. For each dataset, the predictive $p$-value starts to move away from its constant value as data that are incompatible with the model are included in the analysis. That is, the $p_{m_0}$ value begins indicating a change in the level of surprise in the observed data, as soon as the incompatible data are included. As such, based on these visual diagnostics, the chosen threshold would be the lowest threshold such  that there is little change in the constant behaviour of $p_{m_0}$. This clearly occurs at the true threshold of $u=20$ 
for the observed datasets.

The third row of Figure \ref{fig:univariate-simulated} similarly displays boxplots of the posterior predictive $p$-value, $p_{m_0}$, but based on the test statistic $T(y)=q_n(y)=\max\{y_i\}$, the largest observed datapoint. Here, measures of surprise close to $0$ or $1$ would indicate surprise in the degree of compatibility between observed data and model, corresponding to the predictive distribution respectively exhibiting heavier or lighter tails than the observed data. As with the reciprocal likelihood test statistic, the estimated $p_{m_0}$ values are approximately constant around 0.5 at and above the true threshold values, and gradually approach $0$ as the threshold becomes small. Using an empirical quantile as the test statistic more precisely focuses on a single point in the posterior predictive distribution, rather than an overall measure of the full distribution as with the reciprocal likelihood.

The fourth row of Figure \ref{fig:univariate-simulated} illustrates the same analysis as the third row, with the difference that the partial posterior predictive $p$-value, $p_{m^*}$ is shown. In this scenario, the value of the test statistic, $T(t)=\max\{y_i\}$, is excluded when constructing the posterior in each case, in order to avoid double-usage of the data when computing the predictive $p$-value (Section \ref{section:2.1}). The behaviour and interpretation of the $p_{m^*}$ value plot is the same as before, however the minor fluctuations in $p_{m^*}$ are more exaggerated in comparison to $p_{m_0}$ (third row of Figure \ref{fig:univariate-simulated}). For example, there now appears to be a small step-change in the value of $p_{m^*}$ at $u=30$ for the first dataset with $\xi>0$ (bottom-left plot), and a slight upward trend in the bottom-right plot). 
On the one hand, these more exaggerated fluctuations highlight potentially informative features of the observed data for which the Pareto distribution is almost always an approximating model. On the other hand, the extra level of detail may hinder identification of a suitable threshold for further analysis, particularly in situations where the changepoint between the Pareto tail and the remaining body of the observed data is less clear
than the examples considered here.

Figure \ref{Fig:univariate-repeated} shows the same plots as Figure \ref{fig:univariate-simulated}, but based on 50 replicate observed datasets, $y_{obs}$, drawn from the known models. Here, the location and variability of the distribution of the $p$-values is strongly indicative of their performance in identifying the known threshold of $u=20$ for any specific dataset. In each case, there is a clear location change in the $p$-values as the threshold moves below the true threshold. This is quite abrupt for the datasets with very obvious step changes in the observed data (first two columns) and more gradual, though still clear, for the more realistic dataset (right column). Also notable is the variability of the measures of surprise -- for some datasets, the $p$-values can be far from 0.5 when using highly variable test statistics such as $T(y)=\max\{y_i\}$, although this is less apparent for $T(y)=1/f(y|\theta)$. Overall, it is clear that the measures of surprise perform well on average.

Figure \ref{Fig:test_quantiles} (top panels) illustrates the partial posterior predictive $p$-values, $p_{m^*}$, for the first dataset with $\xi>0$, with a range of empirical quantiles, $T(y)=q_j(y)$, as the test statistic. As before, using a test statistic that focuses on a single point of the posterior predictive distribution, produces more variable surprise estimates. This variability occurs as, while there is at least some degree of stability to the posterior predictive distributions for thresholds above $u=20$, the actual value of $p_{m^*}$ is highly dependent on a single point in the observed dataset, which can fluctuate less smoothly with $u$ than the posterior predictive distribution. While it is still possible to determine the true threshold of $u=20$ from these plots, these results suggest that in the absence of any particular need to examine the predictive ability of the model fit at a given point, it may be more practical to use test statistics that compute an overall model fit (such as $T(y)=1/f(y|\theta)$).
The bottom panels in Figure \ref{Fig:test_quantiles} show the variability in the $p$-values obtained by averaging over 50 observed replicate datasets. As for Figure \ref{Fig:univariate-repeated}, the measures of surprise identify the true threshold well on average.

\begin{table} \begin{center} \begin{tabular} { c  c c c c cc}
\hline Dataset & Surprise & $W^2$ & $A^2$ & MRL Plot & Mixture Model & True Mixture Model \\ \hline
$\xi>0$ & 20 & 20 & 20 & 20 & 31.95  (30.48, 34.65) & 19.97 (19.74, 20.09)\\
$\xi<0$ & 20 & 20 & 20 & 20 & 35.83 (34.30, 37.47) & 19.94 (19.73, 20.04)\\
$\xi>0$ & 20  & 20 & 20 & 18 & 27.47 (26.53, 27.92) & 19.90 (18.60, 21.44) \\ \hline
\end{tabular} \caption{\small Comparison of univariate threshold estimates for the three datasets 
 using the measure of surprise,  Cram\'er-von-Mises ($W^2$) and Anderson-Darling ($A^2$) goodness of fit tests, mean-residual life (MRL) plots,  the Bayesian mixture model of MacDonald et al. (2011) and the true mixture model.
 For the mixture model, the values indicate the posterior mean and the 95\% central credibility interval of the threshold estimate.
}\label{Table:U} 
\end{center} \end{table}

Finally, Table \ref{Table:U} provides a summary of the results of implementing a small selection of the many classical and Bayesian threshold selection methods for univariate data, reported to the nearest integer (except for the mixture models). The results based on surprise are those described above. The Cram\'er-von-Mises ($W^2$) and Anderson-Darling ($A^2$) methods (\citeNP{AndersonDarling1954,Stephens1977}) are non-parametric goodness of fit tests comparing the empirical distribution function with the generalised Pareto distribution, $F(y|\xi,\sigma,u)$ (equation \ref{eqn:GPD}). The threshold reported corresponds to the largest (classical) $p$-value not rejecting the Pareto null hypothesis. The mean residual life plot \cite{coles01} identifies the lowest threshold at which the graphical diagnostic becomes linear.
The Bayesian mixture model refers to the method of \shortciteN{macDonald+sldrr11} who fit a mixture of a Pareto distribution above the threshold, $u$, and a mixture of Gaussian distributions (truncated above at $u$) below the threshold. The threshold, $u$, is itself estimated as an unknown parameter. The true mixture model is the same, but fitting the true data-generating model, which is unknown in practice.

The true threshold is correctly identified in most cases,
with the exception of the mixture model, in which the true threshold is not within the estimated 95\% credible interval. (This is a well known phenomena for mixture models when the below-threshold model is not flexible enough to model the non-extreme data, \shortciteNP{macDonald+sldrr11}, and where any threshold $u'>u$ also specifies a valid Pareto tail model; here it is difficult for a mixture of normal distributions to model a uniformly distributed data well.) This is not surprising: the true changepoint is fairly evident in the observed data, although less so in the third dataset, and it is relatively simple to detect this in the univariate setting. However, virtually no methods are available to determine the threshold, $r_0$, in the multivariate setting (see Section \ref{section:intro}). It is in this scenario that the benefits of our approach can be more clearly seen.

\subsection{Multivariate analyses}%

We generated two observed bivariate datasets directly in pseudo-polar co-ordinate form, $(w_{obs},r_{obs})=[(r_{obs,1},w_{obs,1}),\ldots,(r_{obs,n},w_{obs,n})]$,  where in each case $r_{obs,i} \sim f(\xi=0.4,\sigma=10,u=0)$ for $i=1,\ldots,n=3,000$. The angular components are generated from the mixture distributions:
\begin{itemize}

\item Logistic data:
\begin{eqnarray*}
w_{obs,i}|r_{obs,i} & \sim&  p(r_{obs,i})h_L(w|\phi=0.55) + (1-p(r_{obs,i}))h_L(w|\phi=0.3)\\
(r_a,r_b) & = & (21,22)
\end{eqnarray*}

\item Dirichlet mixture data:
\begin{eqnarray*}
w_{obs,i}|r_{obs,i} & \sim&  p(r_{obs,i})h_D^1(w|\mu,\lambda) + (1-p(r_{obs,i}))h_D^2(w|\mu,\lambda)\\
h_D^1(w|\mu,\lambda) & = & 0.25f_D(w|\mu^1=(1,9)) + 0.75f_D(w|\mu^2=(9,1))\\
h_D^2(w|\mu,\lambda) & = & 0.25f_D(w|\mu^1=(4,6)) + 0.75f_D(w|\mu^2=(7,3))\\
(r_a,r_b) & = & (5,8)
\end{eqnarray*}

\end{itemize}
for $i=1,\ldots,n$, where
\[
	p(r) = \left\{\begin{array}{ccl}1&&r \leq r_a\\
	(r-r_b)/(r_a-r_b)& \mbox{for} &r_a < r < r_b\\
	0 && r \geq r_b
\end{array}\right.,
\]
denotes a linear mixing function between $r_a$ and $r_b$. Here 
$h_L(w|\phi)=h(w|\phi=\alpha=\beta)$ is the dependence measure function of the bivariate logistic model (c.f. (\ref{eq:bilogistic}) with $\phi=\alpha=\beta$), and $h_D^1(w|\mu,\lambda)$ and $h_D^2(w|\mu,\lambda)$ are two-component mixture of Dirichlet distributions (\ref{eq:dirichlet}). 
In this manner, each dataset smooth\textcolor{red}{l}y moves from its non-extreme component ($h_L(w|\phi=0.55)$ and $h^1_D(w|\mu,\lambda$)) to its extreme component ($h_L(w|\phi=0.3)$ and $h^2_D(w|\mu,\lambda)$), with the transition complete at $r=r_b$.
In this manner, the true threshold can be considered to be $r_0=r_b$ in each case.

\begin{figure} \begin{center}  
\includegraphics[width=17cm]{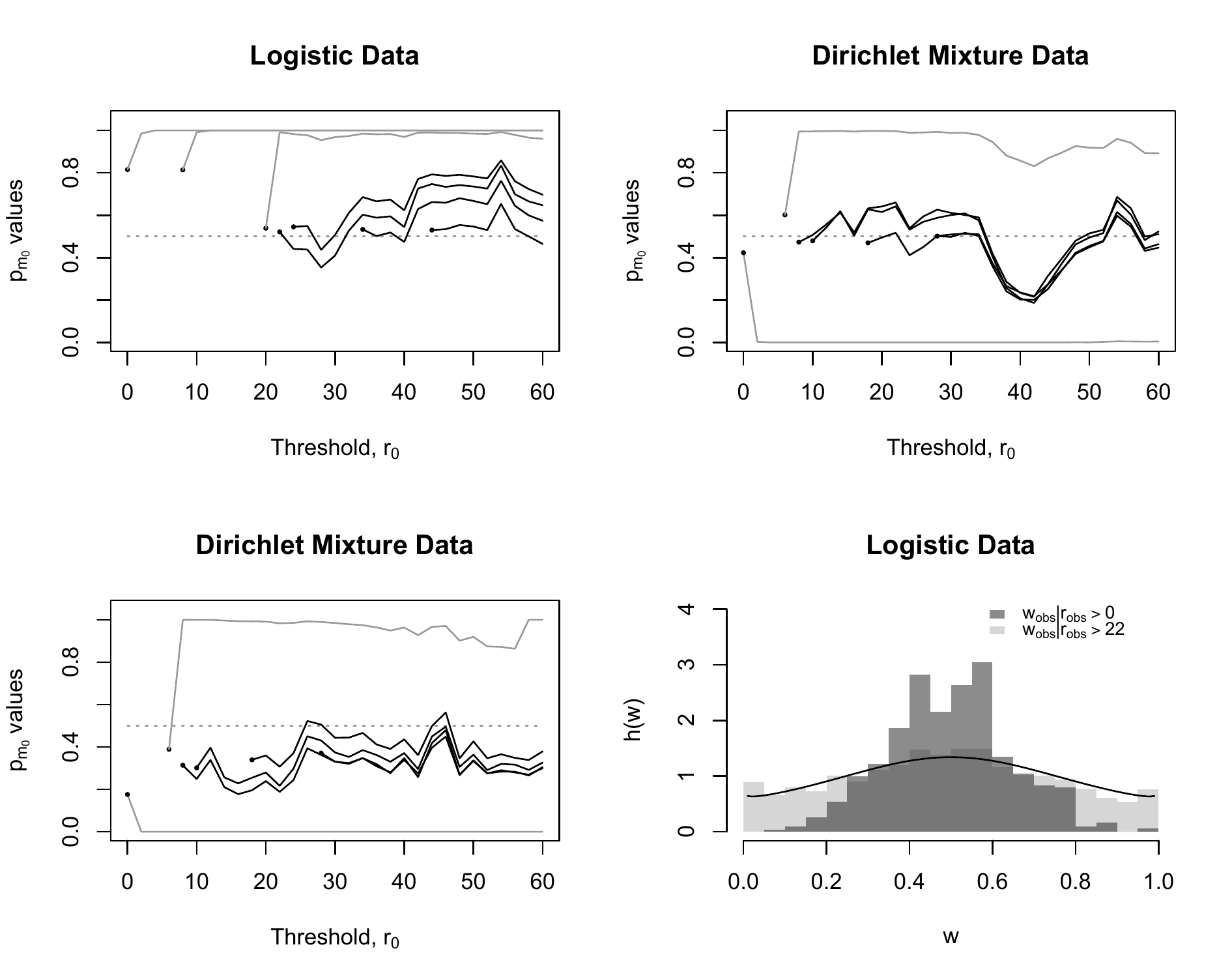}
\end{center}
 \caption{\small
 Measures of surprise for the bivariate Logistic and Dirichlet mixture datasets, as a function of radial threshold $r_0$. Each line indicates the estimate of $p_{m_0}$ at the given threshold, when the posterior is estimated at the leftmost point in each line (shown by a dot). Solid black and grey lines indicate thresholds that are respectively above/at and below the threshold changepoint of $r_0=r_b$.
 Top panels show  when the data-generating model is used to compute $p_{m_0}$. Bottom-left panel uses the logistic model to analyse the Dirichlet mixture dataset.
 Bottom-right plot shows the empirical density of $w_{obs}|r_{obs}>0$ (dark grey bars) and $w_{obs}|r_{obs}>r_b=22$ (light grey) for the Logistic dataset. The solid line illustrates the estimated posterior predictive distribution of the logistic model fitted to $w_{obs}|r_{obs}>0$.
\label{fig:simulated-multivariate}}
 \end{figure}

Figure \ref{fig:simulated-multivariate} shows the estimates of the posterior predictive $p$-value, $p_{m_0}$, when $T(y)=1/f(y|\theta)$, based on uniform priors for all dependence model parameters.  Each line represents the estimate of $p_{m_0}$ at each threshold 
when using the posterior obtained at the leftmost point (indicated by a dot) which also denotes $r_0$ in each case. Solid black and grey lines respectively represent measures of surprise estimated above/at and below the true threshold of $r_0=r_b$.

The top panels in Figure  \ref{fig:simulated-multivariate}
illustrate the surprise estimates when the model used to generate the observed data is also used to construct $p_{m_0}$ in each case. For both analyses, when $r_0$ is specified at or above the true value of $r_b$, the resulting line of $p$-values is approximately constant and centred at 0.5. (The dip in the line of $p$-values for the Dirichlet mixture dataset is a function of the particular dataset used.)
This means that for the posterior evaluated using those $w|r>r_0$ where $r_0$ is given by the dot at the far left of each line of $p$-values, there is no obvious change in the distribution of $w|r>r'$ where $r'\geq r_0$. That is, the data and the model are compatible at $r_0$, and the distribution of $w|r>r'$ is independent of $r'$ for $r'>r_0$, as required by the spectral intensity function (\ref{eq:intensity}) for a suitably high radial threshold. When $r_0$ is below the true value of $r_b$, while the model and data $w|r>r_0$ can appear compatible (that is, the dot can appear close to 0.5), the associated line of $p$-values rapidly approaches 0 or 1, indicating a strong dependence between $w$ and $r$ for the given $r_0$.

The possibility that $p_{m_0}$ can approach 1, indicating that the observed data is apparently {\em more} likely to arise from the model than data generated directly from the model, is an artefact of the test statistic, $T(y)=1/f(y|\theta)$. The bottom-right panel in Figure  \ref{fig:simulated-multivariate} shows histograms of the observed data $w_{obs}|r_{obs}>0$ (i.e. just $w_{obs}$) (dark grey bars) and $w_{obs}|r_{obs}>r_b=22$ (light grey). The black line indicates the estimated posterior predictive distribution of the logistic model fitted to $w_{obs}|r_{obs}>0$. 
 As the observed data for $r_{obs}>0$ matches the predictive distribution passably well, the resulting $p_{m_0}$ value will be not too far from 0.5 (as is observed by the $p_{m_0}=0.8$ dot at $r_0=0$ in Figure \ref{fig:simulated-multivariate}, top-left panel). However, while the observed data $w_{obs}|r_{obs}>r_b=22$ (light grey bars) does not match this predictive distribution well, as the data clusters around the mode of the predictive distribution, the statistic $T(y)=1/f(y|\theta)$ will produce a smaller value than data actually generated from the predictive distribution (and so $p_{m_0}\rightarrow 1$). This feature can be removed through an alternative form of test statistic, although it does not present an obstacle to correct inference in this setting.

However, unlike models for univariate extremes, there is no single, unique parametric limiting distribution for the angular component, $w$, in the multivariate setting. As computation of any measure of surprise requires a specified extremal model (i.e. a given null hypothesis), in practice it is quite likely that the chosen model will not perfectly match the data-generating mechanism of the observed data. The bottom-left panel in Figure \ref{fig:simulated-multivariate} illustrates the outcome when the logistic model is fitted to the Dirichlet mixture data. The interpreted outcome is the same as before, in that the $p$-value lines above the true threshold are approximately constant, and those below the curve quickly drop to zero, thereby easily identifying $r_0=8$ even in the case of a mis-specified model. However, the nature of the model mis-specification,
results in the estimates of $p_{m_0}$ levelling off at values other than 0.5. This occurs as the posterior predictive distribution of the data under the model differs from that of the observed data, and so $p_{m_0}$ may or may not occur at the value 0.5. However, regardless of the model mis-specification, the posterior predictive distribution is robust and does not change for all $r>r_0$, and so the estimates of $p_{m_0}$ remain constant above $r_0$.

As such, the proper interpretation of the bottom-left panel of Figure  \ref{fig:simulated-multivariate} is that due to the approximately constant estimates of $p_{m_0}$, $r_0=8$ may correctly be chosen as the identified threshold for the observed data. However, the estimates of $p_{m_0}$ levelling off at values other than 0.5 indicates that there is an extremal model mis-specification. Of course, in the extreme, where the estimates of $p_{m_0}$ level off at either 0 or 1, it will be impossible to identify a suitable threshold due to the wholly inappropriate null model specification.

\begin{figure} \begin{center}  
\includegraphics[width=16.0cm]{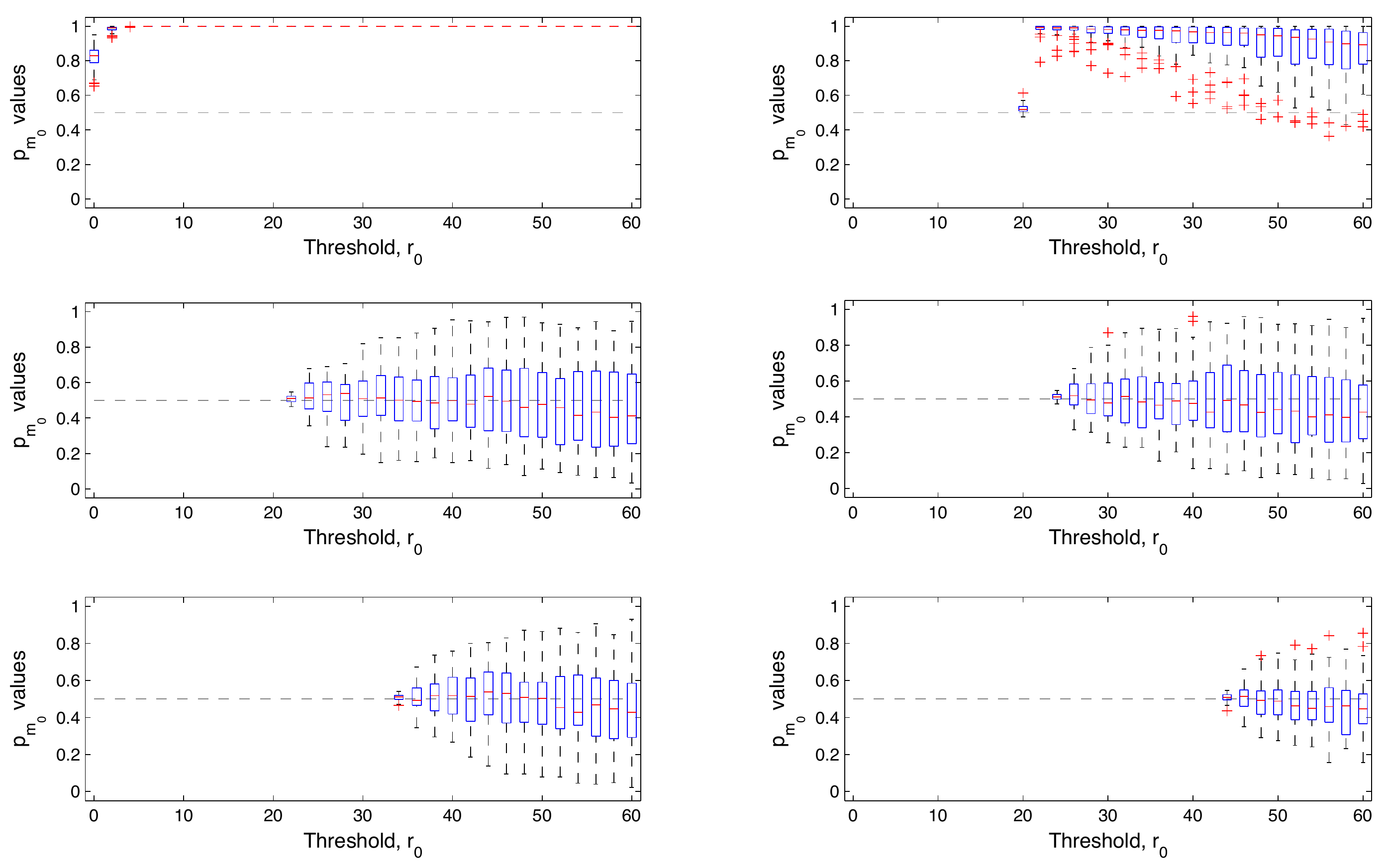}
\end{center}
 \caption{\small
Summaries of the lines of $p_{m_0}$ values in the top-left panel of Figure  \ref{fig:simulated-multivariate}, for the Logistic dataset and model.
Each panel illustrates the variability in $p_{m_0}$ value lines, obtained over 100 replicate observed datasets, for a specific candidate value of $r_0$. Each boxplot corresponds to the range of $p_{m_0}$ values obtained over the replicate datasets, when the candidate value of $r_0$ is given by the location of the lowest boxplot.
The candidate values of $r_0$ correspond to (top-left to bottom-right) 0, 20, 22, 24, 34 and 44. 
The true threshold is at $r_b=22$.
\label{fig:simulated-multivariate-logrep}}
 \end{figure}

\begin{figure} \begin{center}  
\includegraphics[width=16.0cm]{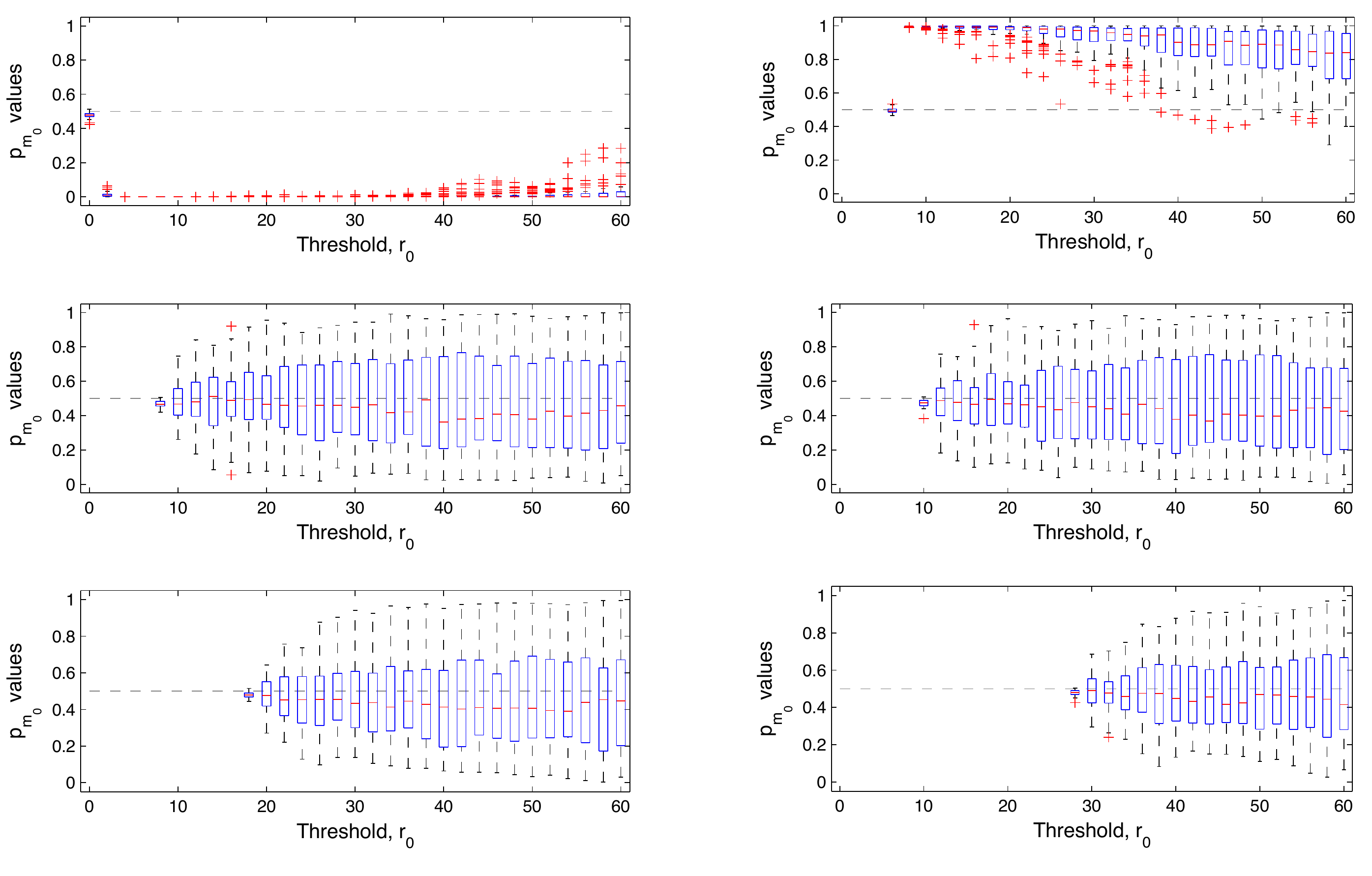}
\end{center}
 \caption{\small
Summaries of the lines of $p_{m_0}$ values in the top-right panel of Figure  \ref{fig:simulated-multivariate}, for the Dirichlet dataset and model.
Each panel illustrates the variability in $p_{m_0}$ value lines, obtained over 100 replicate observed datasets, for a specific candidate value of $r_0$. Each boxplot corresponds to the range of $p_{m_0}$ values obtained over the replicate datasets, when the candidate value of $r_0$ is given by the location of the lowest boxplot.
The candidate values of $r_0$ correspond to (top-left to bottom-right) 0, 6, 8, 10, 18 and 28. 
The true threshold is at $r_b=8$.
\label{fig:simulated-multivariate-dirrep}}
 \end{figure}

Figure \ref{fig:simulated-multivariate-logrep} examines the variability in the lines of $p_{m_0}$ values seen in the top-left panel of Figure \ref{fig:simulated-multivariate} (for the logistic dataset and model), obtained over 100 replicate observed datasets. Each of the replicate lines of $p_{m_0}$ values that start from 0 (i.e. with a candidate value of $r_0$ of 0) are illustrated in the top-left panel of Figure \ref{fig:simulated-multivariate-logrep}, so that any one line of $p_{m_0}$ values takes one value from each boxplot. The panels correspond to the candidate value of $r_0$ being (top-left to bottom-right) 0, 20, 22, 24, 34 and 44. 
When the candidate value of $r_0$ is equal to the true threshold of $r_b=22$ or higher, there is a clear centring of all $p_{m_0}$ values around 0.5 for all thresholds above the candidate threshold. However, when the candidate value of $r_0$ is less than $r_b=22$, then the $p_{m_0}$ values display a markedly different behaviour by rapidly approaching 1. 
The same behaviour is observed in Figure \ref{fig:simulated-multivariate-dirrep}, which displays the same information as Figure \ref{fig:simulated-multivariate-logrep}, but for the Dirichlet mixture data and model (i.e. the top-right panel of Figure \ref{fig:simulated-multivariate}). The true threshold of $r_b=8$ is easily discernible through the clearly different behaviour of the $p_{m_0}$ values at and below the true threshold.
These results indicate that the measures of surprise perform well on average.

\section{Applications}
\label{sec:application}

We now implement the measure of surprise as a tool to determine appropriate thresholds in three real data analyses. In increasing order of data dimensionality, these are the univariate Danish fire loss data \cite{McNeil1997}, a bivariate surge and wave height oceanographic dataset \cite{coles+t94}, and a 5-dimensional, multivariate air quality dataset \cite{HeffernanTawn2004,BoldiDavison2007}.

\subsection{Danish fire loss data}

The Danish fire loss dataset consist of 2,156 losses exceeding one million Danish kroner (DKK) between the years 1980--1990, adjusted to 1985 levels. This dataset has previously been used in the context of threshold identification, with the identified thresholds listed in Table \ref{table:danish}.
\shortciteN{McNeil1997} used a mean residual life plot to determine a reasonably clear appropriate threshold at $u=10$, and perhaps a second at $u=20$. \shortciteN{frigessi+hr03} and \citeN{CabrasCastellanos2009a} implemented various forms of mixture model to identify a somewhat lower threshold of around $u=6$--$8$. An even lower threshold of $u=5.08$ is obtained by  the mixture model approach of \shortciteN{macDonald+sldrr11}.

\begin{figure} \begin{center}  
\includegraphics[width=16cm]{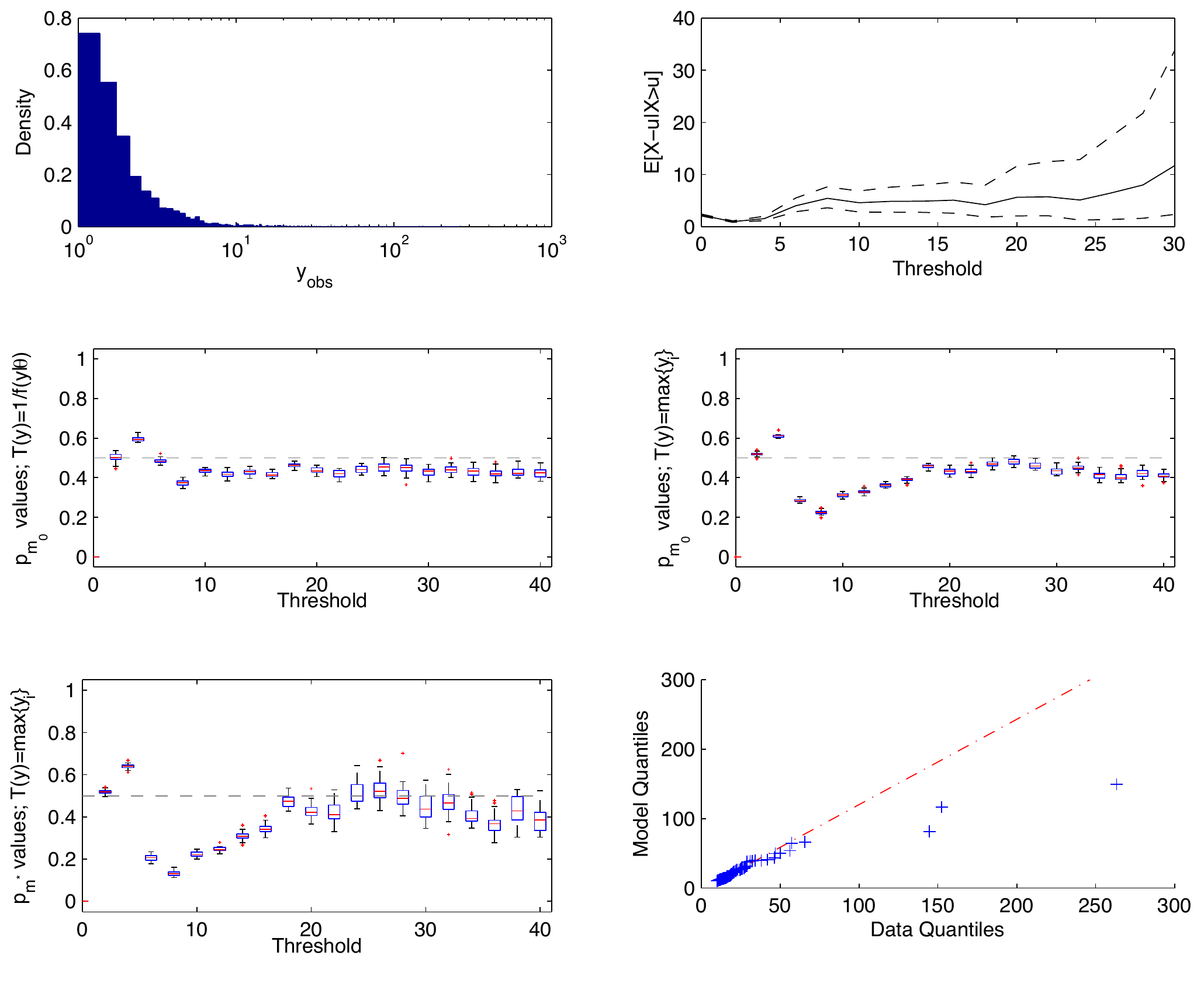}
\end{center}
 \caption{\small
 Measures of surprise for the Danish fire loss dataset, as a function of threshold $u$. Top panels show histogram of data (left) and mean residual life plot (right). Remaining panels illustrate posterior predictive $p$-values,  $p_{m_0}$,  with $T(y)=1/f(y|\theta)$ (top right), and $T(y)=\max\{y_i\}$ (top left), and partial posterior predictive $p$-values, $p_{m^*}$, with $T(y)=\max\{y_i\}$ (bottom left). Bottom right illustrates a quantile-quantile plot of the data for $u>10$ under the Pareto model.
\label{fig:danish}}
 \end{figure}

Figure \ref{fig:danish} illustrates the surprise estimates as a function of $u$, with the centre panels showing the posterior predictive $p$-values, $p_{m_0}$, with $T(\theta)=1/f(y|\theta)$ (left plot) and $T(y)=\max\{y_i\}$ (right plot), and the bottom left panel showing the partial posterior predictive $p$-value, $p_{m^*}$, also with $T(y)=\max\{y_i\}$. Clearly, all three plots show a directional change in the $p$-values below $u=8$, suggesting there is a very clear change in the behaviour of the data at this point. However, both $p_{m_0}$ and $p_{m^*}$ with $T(y)=\max\{y_i\}$, which are focusing on one particular aspect of the dataset, both identify a deviation away from constant behaviour below $u=18$. This  additional change is not apparent when using $T(y)=1/f(y|\theta)$, which evaluates an overall measure of adherence of data and model.

From these results we would most closely agree with the graphical mean residual life graphical analysis of \citeN{McNeil1997} (see top right panel of Figure  \ref{fig:danish}) --  there are indeed two locations (around $u=8$ and $u=18$) where the deviation of the data from the Pareto distribution become most apparent. Of course methods that can only identify a single location (such as mixture models) are less likely to pick this up.

Finally, we comment that the posterior predictive $p$-values appear to become constant at a value just below 0.5, suggesting that the Pareto distribution is in fact not the limiting model describing the tail behaviour of these data. An explanation of this can be obtained by examining the quantile-quantile plot (Figure \ref{fig:danish}, bottom right panel) when $u=10$, and noting that the three largest observations are obviously smaller than expected by the fitted Pareto tail, causing an apparent Pareto model mis-match, which is observed through the measures of surprise.

\begin{table}\begin{center} \begin{tabular} {rc}
\hline  Approach & Threshold, $u$ \\ \hline
\citeN{McNeil1997} & 10 and 20 \\
\shortciteN{frigessi+hr03}  & 6.5 \\
\citeN{CabrasCastellanos2009a} & 5.29 (median), 7.48 (mean) \\ 
\shortciteN{macDonald+sldrr11} & 5.08 (mean),  $95\%$ CI = (4.61,5.29)\\ 
Measure of Surprise & 8 and 18 \\
\hline
\end{tabular} \caption{\small Threshold point estimates for the Danish fire loss dataset, resulting from different threshold identification methods. These include mean residual life plots (McNeil, 1997), various mixture-based approaches (Frigessi et al. 2003; Cabras and Castellanos 2009; MacDonald et al. 2011), a graphical mean residual life plot, and the proposed measure of surprise.
\label{table:danish}}
\end{center} \end{table}

\subsection{Oceanographic data}%

The data consist of a
sequence of hourly coastal surge records for the years 1971-1977 from the port of Newlyn, Cornwell, and corresponding 3-hourly wave records from a light vessel approximately 34.6km offshore. 
These data were previously analysed by \citeN{coles+t94} (see also \citeNP{pugh+v79,smith84}), who produced a bivariate series of $n=2,894$ observations, following a preliminary analysis.
We follow \citeN{coles+t94} in transforming the data to have unit Fr\'echet margins, and 
assume the data are serially independent.
Based on a visual inspection of the histograms of $w_{obs}|r_{obs}>r'$ for a range of values of $r'$,
\citeN{coles+t94} identified $r_0=e^{3.3}=27.11$ as the point above which the histogram of $w_{obs}|r_{obs}>r_0$ did not noticeably change shape.  (\citeNP{coles+t94} actually identified the threshold $r_0=e^{3.3}/n$, however we do not scale the data with $1/n$ in our analysis.) We re-evaluate this choice here.

\begin{figure} \begin{center}  
\includegraphics[width=17cm]{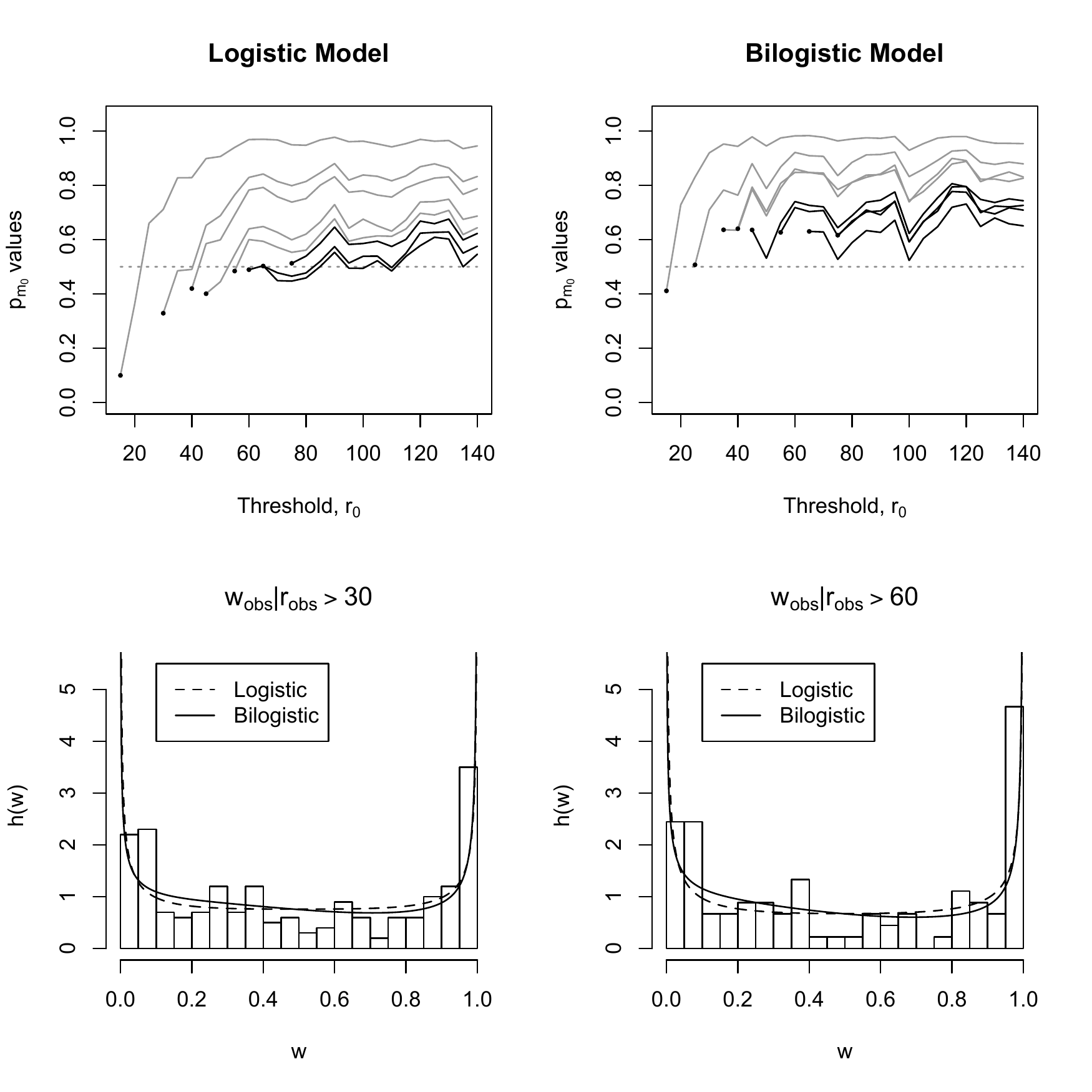}
\end{center}
 \caption{\small
 Measures of surprise for the bivariate oceanographic dataset, as a function of radial threshold $r_0$. Top panels illustrate posterior predictive $p$-values,  $p_{m_0}$,  with $T(y)=1/f(y|\theta)$, using the logistic (left panel) and bilogistic (right panel) models. 
 Each line indicates the estimate of $p_{m_0}$ at the given threshold, when the posterior is estimated at the leftmost point in each line (shown by a dot). 
Grey and black lines indicate values of $p_{m_0}$ below and above the chosen threshold, $r_0$, respectively.
Bottom panels show the histogram of $w_{obs}|r_{obs}>r'$ with $r'=30$ and $60$. Superimposed are the predictive densities of $w$ based on the data $w_{obs}|r_{obs}>r'$, for each extremal model.
\label{fig:ocean}}
 \end{figure}

Figure \ref{fig:ocean} (top panels) illustrates the posterior predictive $p$-value estimates, $p_{m_0}$, with $T(y)=1/f(y|\theta)$, for a range of radial thresholds, $r_0$,
for both logistic and bilogistic models.
For each model it is clear that the threshold of $r_0=27.11$ is too low, as in each case, the line of $p$-values (starting with the dots at $r_0=27.11$) increases rapidly before levelling out at higher thresholds. It is also apparent that the lines of $p$-values do not level off at constant values around 0.5, indicating that neither logistic nor bilogistic model seems to fit the observed data well. For the logistic model, the $p$-value lines seem to continue to slowly climb regardless of the choice of $r_0$, although perhaps for $r_0\approx60$  there is some evidence to suggest that a semi-reasonable threshold has been identified. The situation is clearer for the more flexible bilogistic model, whereby the lines of $p$-values become effectively constant for thresholds greater than $r_0=45$.

The bottom panels in Figure \ref{fig:ocean} present histograms of $w_{obs}|r_{obs}>r'$ for $r'=30$ and $60$, with $r'=30$ corresponding to the radial threshold choice of \citeN{coles+t94}, and the latter, the greater of the two thresholds identified using the measure of surprise. Visually, the two histograms are very similar, although for $r'=60$ there is a slightly lower proportion of data in the range (0.4,0.95). The posterior predictive distributions of each model are superimposed, indicating that the asymmetric bilogistic model is perhaps more suitable than the logistic model. However, there is little obvious difference between the distributions at the two different thresholds.

Hence, it would be very difficult to argue that $r_0=60$ is more appropriate than $r_0=30$ through inspection of the histograms of $w_{obs}|r_{obs}>r'$ alone. It is only through more sophisticated modelling (c.f. Figure \ref{fig:ocean}, top plots) that discrimination between these thresholds becomes possible.

\subsection{Air quality data}

The air quality monitoring data comprise a five-dimensional series of measurements of ground level ozone levels ($\mbox{O}_3$), nitrogen oxide ($\mbox{NO}$), nitrogen dioxide ($\mbox{NO}_2$),  sulphur dioxide ($\mbox{SO}_2$) and particulate matter ($\mbox{PM}_{10}$) in the city centre of Leeds, UK, from 1994-1998. The gases are recorded in parts per billion, and the particulate matter in micrograms per cubic metre. We follow the previous analyses of \citeN{HeffernanTawn2004} and \citeN{BoldiDavison2007} and analyse the data recorded in the winter months of November--February which are assumed to be stationary, and where an entire day is deleted if at least one measurement is missing.  \citeN{BoldiDavison2007} analyse these data using a mixture of Dirichlet distributions (\ref{eq:dirichlet}), 
assuming a threshold of $e^{2.5}\approx 12.18$ based on an {\em ad hoc} method, thereby identifying 46\% of the data as extremes.

\begin{figure} \begin{center}  
\includegraphics[width=17cm]{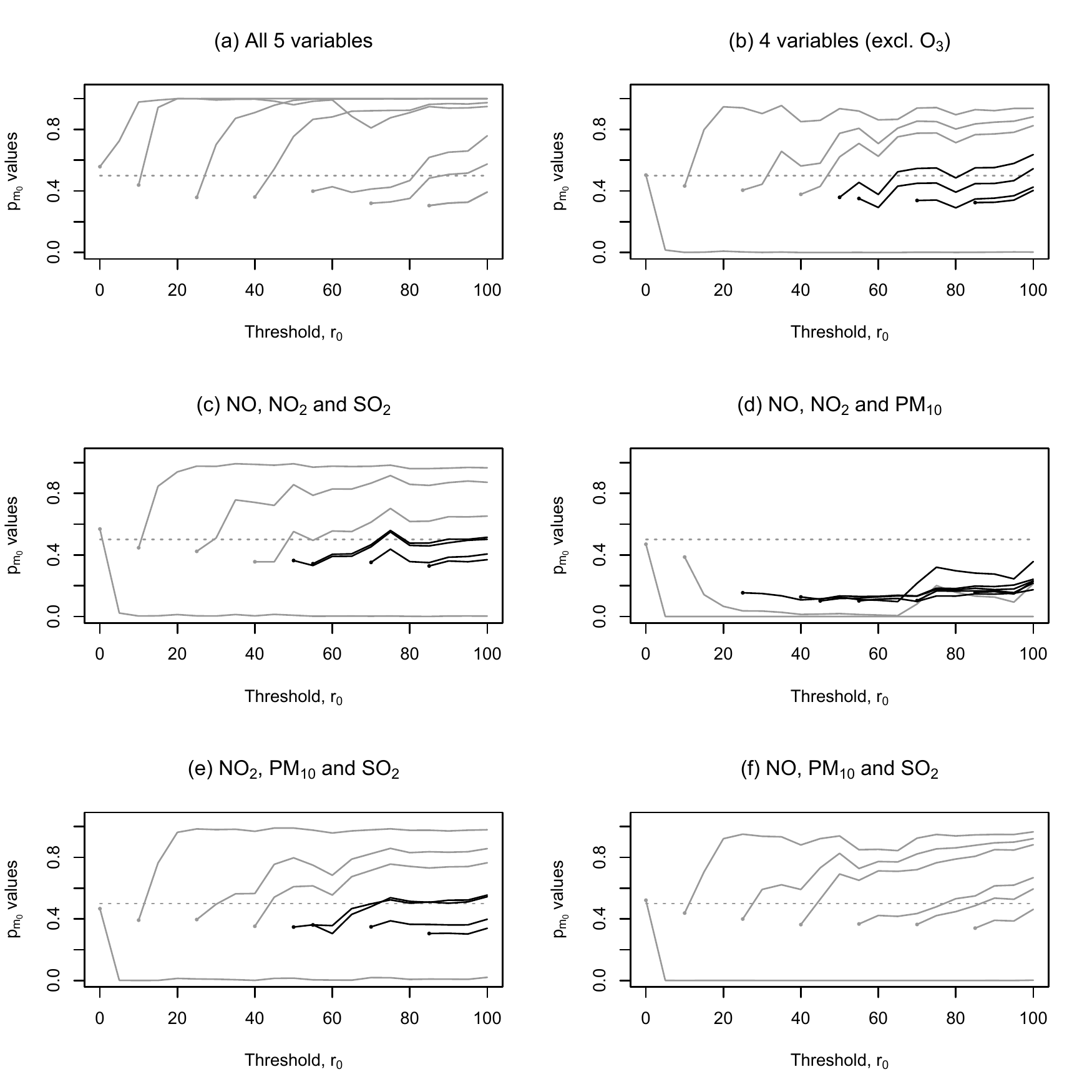}
\end{center}
 \caption{\small 
 Measures of surprise for the 5-dimensional air quality dataset, as a function of radial threshold $r_0$.  Panels illustrate posterior predictive $p$-values,  $p_{m_0}$,  with $T(y)=1/f(y|\theta)$, for various combinations of the 5 variables, using a 2-component mixture of Dirichlet distributions (\ref{eq:dirichlet}). Panel (d) uses a single
 Dirichlet distribution. Each line indicates the estimate of $p_{m_0}$ at the given threshold, when the posterior is estimated at the leftmost point in each line (shown by a dot). 
Grey and black lines indicate values of $p_{m_0}$ below and above the chosen threshold, $r_0$, respectively.
\label{fig:aq}}
 \end{figure}

Figure \ref{fig:aq} illustrates the posterior predictive $p$-value, $p_{m_0}$, for various subsets of the air quality dataset, based on an $I=2$ component mixture of Dirichlet distributions model, and with $T(y)=1/f(y|\theta)$. Panel (a) considers a suitable threshold, $r_0$, for all 5 variables jointly. For $r_0\leq 40$ it is clear that the 2-component mixture of Dirichlet distributions model is inadequate to model the data extremes. The posterior predictive $p$-value rapidly becomes one, and so it is not possible to determine whether it levels off as the model itself is unable to predict the observed data. For $r_0>40$, while $p_{m_0}$ stays within (0,1), there is no indication that the $p$-values level off, and so we are unable to identify a suitable threshold for these data, under the mixture model.

This outcome illustrates a weakness in using measures of surprise to identify a suitable threshold for extreme analyses. Namely, for an appropriate threshold to be selected, it must be possible for the fitted model to have produced the observed data. If the chosen model is not sufficiently flexible to describe the actual extremal dependence structure, then the predictive $p$-values will rapidly reach $0$ or $1$, and no conclusion will be possible. 
However, this same general issue often arises in many other procedures. I.e. if a model is wrong or an estimator is biased, then direct inferences are often invalid.

Figure \ref{fig:aq} (b) shows the same information for the 4-dimensional analysis when excluding $O_3$.  
Here, the exclusion of $O_3$ allows the model to describe the dependence structure of the data reasonably well, and so the measure of surprise is informative. From this plot it is relatively easy to see that below a threshold of $r_0\approx 50$, the $p_{m_0}$ values do not level off, whereas above this value (illustrated by the black lines) they are roughly constant.

Panels (c)--(f) in Figure \ref{fig:aq} examine the threshold for all 3-dimensional subsets of the environmental variables in panel (b). Panels (c) and (e) exhibit similar characteristics to the 4-dimensional analysis, and so it is possible to confirm $r_0\approx 50$ as an acceptable threshold. Panel (f) differs in that the $p$-values above $r_0=50$ do not appear to level off as they do in panels (b), (c) and (e), and exhibit a behaviour closer in character to the 5-dimensional analysis. As such, it is difficult to choose a suitable threshold for these data.
The analysis of $\mbox{NO}$, $\mbox{NO}_2$ and $\mbox{PM}_{10}$ in panel (d) produces markedly different surprise estimates to the others. Here, the fitted model is a single
Dirichlet distribution, as the two-component mixture overfitted the data, and produced a more erratic version of the plot in panel (d).
Quite clearly, for this specific subset of the air quality variables, a much lower threshold of $r_0\approx 25$ can be identified.

While the original analysis of \citeN{BoldiDavison2007} was primarily focused on developing flexible mixture of Dirichlet distributions to model extremal behaviour, our analysis suggests that their adopted threshold of $e^{2.5}\approx 12.18$ is too low for the two-component mixture of Dirichlet distributions model considered here. Of course, using ad-hoc methods to identify the radial threshold, $r_0$, (such as visually examining the empirical distribution of $w|r>r'$ for varying $r'$, \citeNP{coles+t94}) is difficult in two dimensions. This difficulty can only increase as the number of dimensions gets larger.

\section{Conclusion} 

The identification of a suitable threshold is a critical part of any extreme value analysis. In this article we have proposed the use of measures of surprise for this purpose, evaluated through posterior predictive $p$-values. This approach has a number of advantages over existing threshold selection approaches. Aside from being fully Bayesian, because the measure of surprise can be computed only using the extreme data, this means it can avoid the potentially problematic non-/semi-parametric modelling of non-extreme data. As a result, it can equally be applied to the radial thresholds, $r_0$, of multivariate analyses, as well as the more common univariate Pareto threshold, $u$. To the best of our knowledge, this is the first principled approach for threshold identification for multivariate extremes. As a result, in the real data analyses of Section \ref{sec:application}, we have demonstrated that several previous multivariate analyses have likely adopted radial thresholds that are too low.

However, a limitation of this approach stems from the model-based nature of the posterior predictive $p$-value. This requires that multiple datasets be generated under the fitted extremal model. As there are many possible parameterisations of valid models for multivariate extremes, this means that the chosen model is unlikely to perfectly describe the observed extremal dependence structure. In terms of identifying a suitable threshold this will not in itself be a problem if the model is capable of producing the observed data, so that the posterior predictive $p$-values will not take values of $0$ or $1$. However, if the model is wholly unsuitable for the data, then posterior predictive $p$-values of $0$ or $1$ will be produced, and no inference on the threshold will be possible.
Suitable thresholds may additionally not be identifiable if there is unaccounted for heterogeneity  in the data, or if the limiting extremal model is asymptotically dependent when the underlying data generation mechanism is asymptotically independent (and vice versa).

The same criticism can be raised when the analyst is interested in fitting and comparing several multivariate extremal models. As our proposed procedure is model-dependent, this implies that a separate threshold must be identified for each candidate model, and the largest of all thresholds used when comparing the models through e.g. AIC or other criteria. In comparison, the ad-hoc procedure of
\citeN{coles+t94} is able to identify a threshold without any recourse to specific models for $h(w)$ (or $H(w)$), although as we have demonstrated, this approach is unreliable and also becomes rapidly untenable as the model dimension increases.

While fully Bayesian and taking into consideration parameter uncertainty, our approach ultimately selects a fixed threshold. This naturally arises through  the method's construction in avoiding specification of a model for non-extremes. Mixture models for univariate extremes can avoid this and obtain a posterior for $u$, but there are known difficulties in balancing the extreme and non-extreme mixture components \shortcite{macDonald+sl11}. Extending these mixture model ideas to the multivariate setting would appear difficult in comparison to using surprise.

Finally, while we have presented a procedure based on a diagnostic plot, it is credible that an improvement of this approach could develop a more decision-theoretic based approach to threshold identification using measures of surprise, based on the output of multiple posterior predictive distributions (one for each candidate threshold). This would reduce ultimate reliance on correctly visual interpretation of the plot of posterior predictive $p$-values, which is the primary weakness of existing ad-hoc procedures for threshold choice \cite{coles+t94}.

\section*{Acknowledgements}

The authors would like to acknowledge Luis Ra\'ul Pericchi and Irene Vicari for early discussions and investigations in the use of surprise for threshold identification in Pareto models. SAS and YF are also supported by the Australian Research Council (DP0877432 and CE11E0098).

\bibliographystyle{chicago}
\bibliography{threshold}


\end{document}